\documentclass[ALICE,manyauthors]{cernphprep}
\usepackage[comma,square,numbers,sort&compress]{natbib}
\usepackage{hyperref}
\usepackage{lineno}
\usepackage{xspace}
\begin{document}


\newcommand{\CommentBlock}[1]{}


\newcommand{\hT}{\ensuremath{h_{\rm T}}}


\newcommand{\AuAu}{Au+Au}
\newcommand{\PbPb}{Pb--Pb}
\newcommand{\pPb}{p+Pb}
\newcommand{\pbarp}{bar{p}+p}
\newcommand{\pp}{pp}
\newcommand{\pA}{pA}
\newcommand{\aaa}{\ensuremath{A+A}}
\newcommand{\sqrtsNN}{\ensuremath{\sqrt{s_{\mathrm {NN}}}}}
\newcommand{\sqrts}{\ensuremath{\sqrt{s}}}
\newcommand{\rr}{\ensuremath{R}}
\newcommand{\zcut}{\ensuremath{z_{\mathrm{cut}}}}

\newcommand{\intlumi}{\ensuremath{L_{\mathrm{int}}}}
\newcommand{\invmub}{\ensuremath{\mu{b}^{-1}}}
\newcommand{\invnb}{{nb}$^{-1}$}
\newcommand{\gev}{\ensuremath{\mathrm{GeV/}c}}

\newcommand{\Nbin}{\ensuremath{N_{bin}}}
\newcommand{\Npart}{\ensuremath{N_{part}}}
\newcommand{\Nbinavg}{\ensuremath{\left<N_{bin}\right>}}
\newcommand{\TAA}{\ensuremath{T_{AA}}}

\newcommand{\kT}{\ensuremath{k_\mathrm{T}}}
\newcommand{\antikT}{anti-\ensuremath{k_\mathrm{T}}}

\newcommand{\pT}{\ensuremath{p_\mathrm{T}}}
\newcommand{\meanpT}{\ensuremath{\left<p_\mathrm{T}\right>}}

\newcommand{\zg}{\ensuremath{z_\mathrm{g}}}
\newcommand{\nsd}{\ensuremath{n_\mathrm{SD}}}
\newcommand{\Rg}{\ensuremath{R_\mathrm{g}}}

\newcommand{\pTjet}{\ensuremath{p_{\mathrm{T,jet}}}}
\newcommand{\pTjetch}{\ensuremath{p_\mathrm{T,jet}^\mathrm{ch}}}
\newcommand{\pTjetchtrue}{\ensuremath{p_\mathrm{T,jet}^\mathrm{ch,true}}}
\newcommand{\pTjetchsmeared}{\ensuremath{p_\mathrm{T,jet}^\mathrm{ch,smeared}}}
\newcommand{\pTtrig}{\ensuremath{p_{\mathrm{T,trig}}}}
\newcommand{\pTassoc}{\ensuremath{p_{\mathrm{T,assoc}}}}

\newcommand{\pThadron}{\ensuremath{p_{\mathrm{T,hadron}}}}
\newcommand{\pTthresh}{\ensuremath{p_{\mathrm{T,thresh}}}}
\newcommand{\pTh}{\ensuremath{p_{\mathrm{T,h}}}}
\newcommand{\ET}{\ensuremath{E_{\mathrm T}}}

\newcommand{\pizero}{\ensuremath{\pi^0}}
\newcommand{\kzerol}{\ensuremath{\mathrm{K}^{0}_\mathrm{L}}}
\newcommand{\kzeros}{\ensuremath{\mathrm{K}^{0}_\mathrm{S}}}

\newcommand{\zleading}{\ensuremath{z_\mathrm{leading}}}
\newcommand{\CMC}{\ensuremath{C_{\mathrm{MC}}}}
\newcommand{\CMCwithArgs}{\ensuremath{C_{\mathrm{MC}}\left(\pTlow;\pThigh\right)}}
\newcommand{\Acc}{\ensuremath{Acc\left(\pT\right)}}

\newcommand{\DCAxy}{\ensuremath{DCA_{\mathrm xy}}}
\newcommand{\sDCAxy}{\ensuremath{sDCA_{\mathrm xy}}}

\newcommand{\dxidpt}{\ensuremath$\xi=\ln(p_{\rm{T}}^{\rm{jet}}/p_{\rm{T}}^{\rm{track}})$}
\newcommand{\dNdpT}{\ensuremath{\frac{\rm{d}N}{\mathrm{d} \pT}}}
\newcommand{\dNjetdpT}{\ensuremath{\frac{\rm{d}^{2}N^\mathrm{AA}_{jet}}{\mathrm{d}\pTjetch\mathrm{d}\eta_\mathrm{jet}}}}
\newcommand{\dNjetdpTdphi}{\ensuremath{\frac{\rm{d}^{2}N_{jet}}{\mathrm{d}\pTjetch\mathrm{d}\dphi}}}

\newcommand{\dNjet}{\ensuremath{\mathrm{d}N_{\rm{jet}}}}
\newcommand{\dN}{\ensuremath{\mathrm{d}N}}
\newcommand{\dpT}{\ensuremath{\delta{\pT}}}
\newcommand{\dpTRC}{\ensuremath{\delta{p_{T,RC}}}}
\newcommand{\pTembed}{\ensuremath{\pT^{\rm{part,embed}}}}

\newcommand{\dNdxi}{\ensuremath{\mathrm{d}N/\mathrm{d} \xi}}

\newcommand{\qhat}{\ensuremath{\hat{q}}}
\newcommand{\vtwo}{\ensuremath{v_2}}


\newcommand{\Ntrig}{\ensuremath{\mathrm{N}^\mathrm{AA}_{\rm{trig}}}}
\newcommand{\Njets}{\ensuremath{N_{\rm{jets}}}}

\newcommand{\sigr}{\ensuremath{\sigma_r}}
\newcommand{\ra}{\ensuremath{r_a}}
\newcommand{\rb}{\ensuremath{r_b}}
\newcommand{\dab}{\ensuremath{\delta_{ab}}}

\newcommand{\sigra}{\ensuremath{\sigma_{r,a}}}
\newcommand{\sigrb}{\ensuremath{\sigma_{r,b}}}
\newcommand{\sigc}{\ensuremath{\sigma_c}}
\newcommand{\sigdab}{\ensuremath{\sigma_{\delta,ab}}}

\newcommand{\Ntriga}{\ensuremath{N_{\rm{trig,a}}}}
\newcommand{\Ntrigb}{\ensuremath{N_{\rm{trig,b}}}}

\newcommand{\RAA}{\ensuremath{R_\mathrm{AA}}}

\newcommand{\Drecoil}{\ensuremath{\Delta_\mathrm{recoil}}}
\newcommand{\cRef}{\ensuremath{c_\mathrm{Ref}}}

\newcommand{\IAA}{\ensuremath{I_{\rm{AA}}}}
\newcommand{\DIAA}{\ensuremath{\Delta\IAA}}

\newcommand{\SpT}{\ensuremath{S\left(\pT\right)}}

\newcommand{\pTconst}{\ensuremath{p_\mathrm{T,const}}}

\newcommand{\pTpr}{\ensuremath{p_\mathrm{T}^\prime}}

\newcommand{\pTcorr}{\ensuremath{p_\mathrm{T,jet}^\mathrm{corr,ch}}}
\newcommand{\pTcorri}{\ensuremath{p_\mathrm{T,jet}^\mathrm{corr,i}}}

\newcommand{\pTraw}{\ensuremath{p_\mathrm{T,jet}^\mathrm{raw,ch}}}
\newcommand{\pTrawi}{\ensuremath{p_\mathrm{T,jet}^\mathrm{raw,i}}}

\newcommand{\pTreco}{\ensuremath{p_\mathrm{T,jet}^\mathrm{reco,ch}}}
\newcommand{\Ajet}{\ensuremath{A_\mathrm{jet}}}
\newcommand{\rhoA}{\ensuremath{\rho\cdot\Ajet}}

\newcommand{\pTcorrdet}{\ensuremath{p_\mathrm{T}^\mathrm{corr,det}}}
\newcommand{\pTrecodet}{\ensuremath{p_\mathrm{T,jet}^\mathrm{reco,det}}}
\newcommand{\Ajetdet}{\ensuremath{A_\mathrm{jet}^\mathrm{det}}}
\newcommand{\rhoAdet}{\ensuremath{\rho^\mathrm{det}\cdot\Ajetdet}}

\newcommand{\pTcorrpart}{\ensuremath{p_\mathrm{T}^\mathrm{corr,part}}}
\newcommand{\pTrecopart}{\ensuremath{p_\mathrm{T,jet}^\mathrm{reco,part}}}
\newcommand{\Ajetpart}{\ensuremath{A_\mathrm{jet}^\mathrm{part}}}
\newcommand{\rhoApart}{\ensuremath{\rho^\mathrm{part}\cdot\Ajetpart}}

\newcommand{\pTdet}{\ensuremath{p_\mathrm{T,jet}^\mathrm{det}}}
\newcommand{\dpTdet}{\ensuremath{\delta{p}_\mathrm{T,jet}^\mathrm{det}}}
\newcommand{\RdpTdet}{\ensuremath{{\bf R}\left[\delta{p}_\mathrm{T}^\mathrm{det}\right]}}
\newcommand{\RinvdpTdet}{\ensuremath{\widetilde{{\bf R}^{-1}}\left[\delta{p}_\mathrm{T}^\mathrm{det}\right]}}

\newcommand{\RinvJetDetPart}{\ensuremath{\widetilde{{\bf R}^{-1}}\left[p_{T}^{jet,part}\rightarrow p_{T}^{jet,det}\right]}}
\newcommand{\RinvCorrDetPart}{\ensuremath{\widetilde{{\bf R}^{-1}}\left[p_{T}^{corr,part}\rightarrow p_{T}^{corr,det}\right]}}
\newcommand{\RCorrDetPart}{\ensuremath{R\left[p_{T}^{corr,part}\rightarrow p_{T}^{corr,det}\right]}}
\newcommand{\RDetPart}{\ensuremath{{\bf R}\left[\pTpart,\pTdet\right]}}

\newcommand{\RpTprPart}{\ensuremath{{\bf R}\left[\pTpart,\pTpr\right]}}

\newcommand{\RinvDetPart}{\ensuremath{\widetilde{{\bf R}^{-1}}\left[\pTpart,\pTdet\right]}}
\newcommand{\RinvDetPartprecise}{\ensuremath{{\bf R}^{-1}\left[\pTpart,\pTdet\right]}}

\newcommand{\partJetEff}{\ensuremath{\epsilon(\pTpart)}}
\newcommand{\partJetEffprecise}{\ensuremath{\epsilon^\prime(\pTpart)}}

\newcommand{\pTpart}{\ensuremath{p_\mathrm{T,jet}^\mathrm{part}}}
\newcommand{\dpTpart}{\ensuremath{\delta{p}_\mathrm{T,jet}^\mathrm{part}}}
\newcommand{\RdpTpart}{\ensuremath{{{\bf R}\left[\delta{p}_\mathrm{T}^\mathrm{part}\right]}}}
\newcommand{\RinvdpTpart}{\ensuremath{\widetilde{{\bf R}^{-1}}\left[\delta{p}_\mathrm{T}^\mathrm{part}\right]}}

\newcommand{\pTrecoi}{\ensuremath{p_\mathrm{T,jet}^\mathrm{reco,i}}}
\newcommand{\pTi}{\ensuremath{p_\mathrm{T,jet}^\mathrm{i}}}
\newcommand{\Ajeti}{\ensuremath{A_\mathrm{jet}^\mathrm{i}}}

\newcommand{\rhoAi}{\ensuremath{\rho\cdot{A_\mathrm{jet}^\mathrm{i}}}}
\newcommand{\Ai}{\ensuremath{A_i}}

\newcommand{\pTlowRef}{\ensuremath{p_\mathrm{T}^\mathrm{ref,low}}}
\newcommand{\pThighRef}{\ensuremath{p_\mathrm{T}^\mathrm{ref,high}}}
\newcommand{\pTlowSig}{\ensuremath{p_\mathrm{T}^\mathrm{sig,low}}}
\newcommand{\pThighSig}{\ensuremath{p_\mathrm{T}^\mathrm{sig,high}}}

\newcommand{\phiTT}{\ensuremath{\varphi_\mathrm{TT}}}
\newcommand{\phiTrig}{\ensuremath{\varphi_\mathrm{trig}}}
\newcommand{\phiJet}{\ensuremath{\varphi_\mathrm{jet}}}
\newcommand{\dphi}{\ensuremath{\Delta\varphi}}

\newcommand{\TTSig}{\ensuremath{\mathrm{TT}_{\mathrm{Sig}}}}
\newcommand{\TTRef}{\ensuremath{\mathrm{TT}_{\mathrm{Ref}}}}

\newcommand{\dpTbin}{\ensuremath{\Delta{p}_{T,jet}^{bin}}}

\newcommand{\Rinstr}{\ensuremath{R_{\mathrm{instr}}}}
\newcommand{\Rfull}{\ensuremath{R_{\mathrm{full}}}}

\newcommand{\Rdet}{\ensuremath{R_{\mathrm{det}}}}
\newcommand{\Rtot}{\ensuremath{R_{\mathrm{tot}}}}

\newcommand{\pTrec}{\ensuremath{p_{\mathrm{T,jet}}^{\mathrm{det}}}}
\newcommand{\pTgen}{\ensuremath{p_{\mathrm{T,jet}}^{\mathrm{part}}}}
\newcommand{\EffJetFind}{\ensuremath{\epsilon_{\rm{jf}}}}
\newcommand{\EffKin}{\ensuremath{\epsilon_{\rm{kin}}}}

\newcommand{\Rargs}{\ensuremath{\left(\pTrec,\pTgen\right)}}

\newcommand{\Niter}{\ensuremath{N_{iter}}}
\newcommand{\Dphirecoil}{\ensuremath{\Delta\varphi_{recoil}}}
\newcommand{\Drecoilphi}{\ensuremath{\Phi(\dphi)}}

\newcommand{\phithresh}{\ensuremath{\Delta\varphi_{\mathrm{thresh}}}}
\newcommand{\Dyieldthresh}{\ensuremath{\Sigma\left(\phithresh\right)}}

\newcommand{\xvtx}{\ensuremath{x_\mathrm{vtx}}}
\newcommand{\yvtx}{\ensuremath{y_\mathrm{vtx}}}
\newcommand{\zvtx}{\ensuremath{z_\mathrm{vtx}}}
\newcommand{\zvtxSPD}{\ensuremath{z_\mathrm{vtx}^\mathrm{SPD}}}


\newcommand{\pptohjet}{\rm{\pp\rightarrow\rm{h}+{jet}+X}}
\newcommand{\pptoh}{\rm{\pp\rightarrow\rm{h}+X}}

\newcommand{\AAtohjet}{\mathrm{AA}\rightarrow\rm{h}+{jet}+X}
\newcommand{\AAtoh}{\mathrm{AA}\rightarrow\rm{h}+X}

\newcommand{\SumBkgHad}{\sum p_{\rm{T,h}}^{\rm{det,bkg}}}
\newcommand{\SumJetHad}{\sum p_{\rm{T,h}}^{\rm{det,signal}}}
\newcommand{\SumJetPart}{\sum p_{\rm{T,h}}^{\rm{part,signal}}}
\newcommand{\RdeltaPt}{\ensuremath{{\bf R}\left[\delta(\SumBkgHad)\rightarrow 0\right]}}
\newcommand{\RinvdeltaPt}{\ensuremath{\widetilde{{\bf R}^{-1}}\left[\delta(\SumBkgHad)\rightarrow 0\right]}}
\newcommand{\RinvJetDetToPart}{\ensuremath{\widetilde{{\bf R}^{-1}}\left[\SumJetHad\rightarrow\SumJetPart\right]}}

\newcommand{\pTjetfull}{\ensuremath{p_{\rm{T,jet}}^{\rm{full}}}}
\newcommand{\dNdpTch}{\ensuremath{\frac{\rm{dN_{jet}}}{\mathrm{d}\pTjetch}}}
\newcommand{\dNdpTfull}{\ensuremath{\frac{\rm{dN_{jet}}}{\mathrm{d}\pTjetfull}}}
\newcommand{\Rchfull}{\ensuremath{R(\pTjetfull,\pTjetch)}}
\newcommand{\effch}{\ensuremath{\epsilon(\pTjetch)}}


\newcommand{\sumievtijet}{\ensuremath{\sum_{ievt}\ \sum_{ijet\in{ievt}}}}

\newcommand{\BnpT}[2]{\ensuremath{B_{#2}(#1)}}
\newcommand{\Phip}[2]{\ensuremath{\Phi_{#2}(#1)}}

\newcommand{\nijet}{\ensuremath{n(ijet)}}

\newcommand{\Sdet}[2]{\ensuremath{S_{#1}^{det,#2}}}
\newcommand{\DSpartdet}[2]{\ensuremath{{\Delta}S_{#1}^{part\rightarrow{det},#2}}}
\newcommand{\DSdetpart}[2]{\ensuremath{{\Delta}S_{#1}^{det\rightarrow{part},#2}}}

\newcommand{\RdeltaPtpmj}{\ensuremath{{\bf R}\left[0 \rightarrow \dpTdet\right]}}

\newcommand{\DeltaR}{\ensuremath{\Delta{R}}}

\newcommand{\pTcorrdetn}{\ensuremath{p_{\mathrm{T},n}^\mathrm{corr,det}}}

\newcommand{\pTlow}[1]{\ensuremath{p_{{\mathrm T},#1}^{\mathrm{low}}}}
\newcommand{\pThigh}[1]{\ensuremath{p_{{\mathrm T},#1}^{\mathrm{high}}}}

\newcommand{\philow}[1]{\ensuremath{\varphi_{#1}^{\mathrm{low}}}}
\newcommand{\phihigh}[1]{\ensuremath{\varphi_{#1}^{\mathrm{high}}}}
\newcommand{\phijet}{\ensuremath{\varphi_{ijet}}}

\newcommand{\dpTlow}[1]{\ensuremath{{\delta}p_{{\mathrm T},#1}^{\mathrm{low}}}}
\newcommand{\dpThigh}[1]{\ensuremath{{\delta}p_{{\mathrm T},#1}^{\mathrm{high}}}}

\newcommand{\pTcorrdetijet}{\ensuremath{p_{\mathrm{T},ijet}^\mathrm{corr,det}}}

\newcommand{\SumBkgHadijet}{\ensuremath{\sum p_{\rm{T,h,ijet}}^{\rm{det,bkg}}}}
\newcommand{\SumJetHadijet}{\ensuremath{\sum p_{\rm{T,h,ijet}}^{\rm{det,signal}}}}

\newcommand{\rhoAindex}{\ensuremath{\rho^{det,ievt}\cdot{A_{ijet}^{det}}}}

\newcommand{\SumBkgHadmatched}{\ensuremath{\sum p_{\rm{T,h,matched}}^{\rm{det,bkg}}}}
\newcommand{\rhoAmatched}{\ensuremath{\rho^{det,ievt}\cdot{A_{matched}^{det}}}}

\newcommand{\pTtrue}{\ensuremath{p_T^{true}}}
\newcommand{\pTsmear}{\ensuremath{p_T^{smeared}}}

\newcommand{\PldpT}{\ensuremath{P_l^{embed}(\pTpart)}}
\newcommand{\Pldata}{\ensuremath{P_l^{data}(\pT)}}
\newcommand{\dPldata}{\ensuremath{{\delta}P_l^{data,correl}(\pT)}}

\newcommand{\dpTpartbin}[1]{\ensuremath{\delta{p_{T,#1}^{part}}}}

\newcommand{\rmn}[2]{\ensuremath{\left<r\right>^{}_{{#1},{#2}}}}
\newcommand{\rmnp}[3]{\ensuremath{r^{}_{{#1},{#2};{#3}}}}
\newcommand{\cmnp}[3]{\ensuremath{c^{}_{{#1},{#2};{#3}}}}
\newcommand{\drmnp}[3]{\ensuremath{\Delta{r}^{}_{{#1},{#2};{#3}}}}

\newcommand{\rinvmn}[2]{\ensuremath{\left<r\right>^{-1}_{{#1},{#2}}}}
\newcommand{\rinvmnp}[3]{\ensuremath{r^{-1}_{{#1},{#2};{#3}}}}
\newcommand{\cinvmnp}[3]{\ensuremath{c^{-1}_{{#1},{#2};{#3}}}}

\begin{titlepage}
\PHyear{2019}       
\PHnumber{087}      
\PHdate{25 April}  

\title{Exploration of jet substructure using iterative declustering in pp and Pb--Pb collisions at LHC energies}
\ShortTitle{Exploration of jet substructure using iterative declustering}   

\Collaboration{ALICE Collaboration\thanks{See Appendix~\ref{app:collab} for the list of collaboration members}}
\ShortAuthor{ALICE Collaboration} 

\begin{abstract}

The ALICE collaboration at the CERN LHC reports novel measurements of jet substructure in pp collisions at $\sqrts~= 7$ TeV and central Pb--Pb collisions at $\sqrtsNN~= 2.76$~TeV.  Jet substructure of track-based jets is explored via iterative declustering and grooming techniques. We present the measurement of the momentum sharing of two-prong substructure exposed via grooming, the \zg, and its dependence on the opening angle, in both pp and Pb--Pb collisions.  We also present the  measurement of the distribution of the number of branches obtained in the iterative declustering of the jet, which is interpreted as the number of its hard splittings. In Pb--Pb collisions, we observe a suppression of symmetric splittings at large opening angles and an enhancement of splittings at small opening angles relative to pp collisions, with no significant modification of the number of splittings.  
 The results are compared to predictions from various Monte Carlo event generators to test the role of important concepts in the evolution of the jet in the medium such as color coherence. 

\end{abstract}
\end{titlepage}

\setcounter{page}{2} 

\section{Introduction}

The objective of the heavy-ion jet physics program at the LHC is to probe fundamental, microscopic properties of nuclear matter at high densities and temperatures. Jets provide well-calibrated probes of the dense medium created in heavy-ion collisions. In pp collisions, the production of jets and their substructure have been measured extensively and these measurements are well-reproduced by theoretical calculations based on perturbative QCD (pQCD) (see Refs.~\cite{Dasgupta:2016bnd,Asquith:2018igt,Sirunyan:2018asm,Larkoski:2017jix} and citations therein). Jets are produced in high-momentum transfer processes, which occur on time scales much shorter than the formation time of the Quark-Gluon Plasma (QGP) generated in heavy-ion collisions; the production rates of  jets in heavy-ion collisions can therefore be calculated accurately using the same pQCD approaches as for pp collisions, after taking into account the effects of nuclear geometry and nuclear modification of parton distribution functions (PDFs) \cite{Eskola:2010jh}. 

Jets traversing the QGP will interact via elastic and radiative processes which modify the reconstructed jet cross section and structure relative to jets in vacuum (``jet quenching'') \cite{Majumder:2010qh}.
Jet quenching effects have been extensively observed in nuclear collisions at RHIC and LHC in measurements of inclusive production and correlations of high-p$_{\rm T}$ hadrons and jets, including correlations of high-energy triggers (hadrons, photons, W and Z bosons, and jets) and reconstructed jets \cite{Adam:2015doa,Adamczyk:2017yhe,Sirunyan:2017jic,Aaboud:2018anc} as well as in the measurement of jet shapes \cite{Acharya:2018uvf,Chatrchyan:2013kwa,Sirunyan:2018qec,Aaboud:2017bzv,Acharya:2017goa,Sirunyan:2018gct}. Comparisons of these measurements to theoretical jet quenching calculations enable the determination of dynamical properties of the QGP, notably the transport parameter $\hat{q}$ \cite{Burke:2013yra}.

More recently, the modification of the jet substructure due to jet quenching has been explored in heavy-ion collisions using tools developed for the measurement of jet substructure in pp collisions for QCD studies and Beyond Standard Model searches \cite{Bendavid:2018nar,Asquith:2018igt}.  A key tool is iterative declustering, which subdivides jets into branches or splittings that can be projected onto the phase space of such splittings, called the Lund plane~\cite{Andersson:1988gp,Dreyer:2018nbf,Andrews:2018jcm}. While the splitting map contains kinematic information of all splittings, techniques like grooming~\cite{Dasgupta:2013ihk,Larkoski:2014wba} can be applied to isolate a specific region of the splitting map according to different criteria such as mitigation of non-perturbative effects, enhancement of the jet quenching signal or simplification of perturbative calculations. 

In this work we focus on the Mass Drop\cite{Dasgupta:2013ihk} or Soft Drop (SD) grooming~\cite{Larkoski:2014wba} with $z_{\rm{cut}}=0.1$ and $\beta=0$. This technique selects the first splitting in the declustering process for which the subleading prong carries a fraction $z$ of the momentum of the emitting prong larger than $z_{\rm{cut}}$. Note that this criterion selects a subset of the splittings. The grooming procedure removes soft radiation at large angles to expose a two-prong structure in the jet.  The shared momentum fraction of those prongs is called $z_{\rm{g}}$, the groomed subjet momentum balance. The measurement of $\it{z}_{\rm{g}}$ in vacuum is closely related to the Altarelli-Parisi splitting functions \cite{Larkoski:2014wba}. 

Theoretical considerations of the in-medium modification of $z_{\rm g}$ can be found in~\cite{Milhano:2017nzm,DEramo:2018eoy,Mehtar-Tani:2016aco,Chang:2017gkt}. 
A key physics ingredient in the theoretical calculations is color coherence\cite{CasalderreySolana:2012ef}. This is the effect by which a color dipole cannot be resolved by the medium as two independent color charges if the opening angle of the dipole is small compared to a fundamental medium scale. If the dipole cannot be resolved, it will propagate through the medium as a single color charge. If color coherence is at work, there will be parts of the jet substructure that won't be resolved, leading to a reduced effective number of color charges and thus a reduced amount of energy loss in medium. 

With the grooming technique we select a hard two-prong substructure. Then we inspect the dependence on the opening angle of the rate of such two-prong objects in medium relative to vacuum. We are interested in understanding whether large-angle splittings are more suppressed relative to vacuum than small-angle splittings, as one would expect if large-angle splittings are  resolved by the medium and radiate in the medium incoherently. 
Previous measurements of \zg~by the CMS collaboration~\cite{Sirunyan:2017bsd} show a modification in central Pb--Pb collisions relative to the pp reference whilst measurements performed at RHIC by the STAR collaboration showed no modification~\cite{Kauder:2017mhg}. Those measurements did not scan the $\DeltaR$ dependence and cover different intervals of the subleading prong energies that can bias towards different typical splitting formation times.  

This work reports the measurement by the ALICE collaboration of $z_{\rm{g}}$, the shared momentum fraction of two-prong substructure, its dependence on the opening angle and $\it{n}_{\rm{SD}}$, the number of splittings satisfying the grooming condition obtained via the iterative declustering of the jet \cite{Frye:2017yrw}, in pp collisions at $\sqrts~=$ 7 TeV and central (0--10$\%$) Pb--Pb collisions at $\sqrtsNN~=2.76$ TeV. 


\section{Data sets and event selection}

A detailed description of the ALICE detector and its performance can be found in Refs.~\cite{Aamodt:2008zz,Abelev:2014ffa}. The analysed pp data were collected during Run 1 of the LHC in 2010 with a collision centre-of-mass energy of $\sqrts~= 7$ TeV using a minimum bias (MB) trigger. The MB trigger configuration is the same as described in Ref. \cite{ALICE:2014dla}.
The data from heavy-ion collisions were recorded in 2011 at $\sqrtsNN~= 2.76$ TeV. This analysis uses the 0--10\% most-central Pb--Pb collisions selected by the online trigger based on the hit multiplicity measured in the forward V0 detectors~\cite{Abbas:2013taa}.  
The datasets and event selection are identical to Refs. \cite{Adam:2015doa,Acharya:2018uvf}. After offline selection, the pp sample consists of 168 million events, while the Pb--Pb sample consists of 19 million events.

 The analysis uses charged tracks reconstructed by the Inner Tracking System (ITS)~\cite{Aamodt:2010aa} and Time Projection Chamber (TPC)~\cite{2010NIMPA.622..316A} which both cover the full azimuth and pseudo-rapidity $|\eta| < 0.9$.  Tracks are required to have transverse momentum $0.15 <p_{\rm T} <100$ GeV/$c$. The track selection is slightly different in the analysis of the 2010 and the 2011 data. The former uses a subclass of tracks with worse momentum resolution that is excluded from the latter \cite{Adam:2015ewa}.  

In pp collisions, the tracking efficiency is approximately 80\% for tracks with $\pT > 1~\gev$, decreasing to roughly 56\% at $\pT = 0.15~\gev$, with a track momentum resolution of 1\% for $\pT = 1~\gev$ and 4.1\% for $\pT = 40~\gev$~\cite{Abelev:2014ffa,Abelev:2012hxa,ALICE:2014dla}. 
In Pb--Pb collisions, the tracking efficiency is about 2 to 3$\%$ worse than in pp. The track $p_{\rm T}$ resolution is about 1\% at $\pT = 1~\gev$ and 2.5\% for $\pT = 40~\gev$. 

As a vacuum reference for the Pb--Pb measurements we use simulated pp collisions at $\sqrts~=2.76$ TeV, calculated using PYTHIA 6.425 (Perugia Tune 2011) [27] and embedded into real central Pb--Pb events at the detector level, to take into account the smearing by the background fluctuations. 
We use the embedding of PYTHIA-generated events instead of the embedding of real pp data measured at $\sqrts~=2.76$ TeV due to the limited size of the data sample. The PYTHIA MC describes well vacuum intrajet distributions \cite{Asquith:2018igt}. In this paper, we validate the PYTHIA calculation by comparing it to jet substructure measurements in pp collisions at $\sqrts~=7$ TeV. 

\section{Jet reconstruction}

Jets are reconstructed from charged tracks using the anti-$k_{\rm{T}}$ algorithm~\cite{FastJetAntikt} implemented in FastJet~\cite{Cacciari:2011ma} with a jet resolution parameter of
$R=0.4$. The four-momenta of tracks are combined using the E-scheme recombination~\cite{Cacciari:2011ma} where the pion mass is assumed for all reconstructed tracks. In order to ensure that all jet candidates are fully contained within the fiducial volume of the ALICE detector system, accepted jets were required to have their centroid constrained to $|\eta_{\rm{jet}}| <$ 0.5. 

The jet finding efficiency is 100$\%$ in the measured kinematic ranges. The jet energy instrumental resolution is similar for pp  and Pb--Pb collisions, varying from 15\% at \pTjetch~$= 20$ \gev~to 25\% at \pTjetch~$= 100$ \gev. The Jet Energy Scale (JES) uncertainty is dominated by the tracking efficiency uncertainty which is 4$\%$.

In pp collisions, no correction for the underlying event is applied. In Pb--Pb collisions, the jet energy is partially adjusted for the effects of uncorrelated background using the constituent subtraction method~\cite{Berta:2014eza}. Constituent subtraction corrects individual jet constituents by modifying their four-momentum. The momentum that is subtracted from the constituents is determined using the underlying event density, $\rho$, which is calculated by clustering the event into $R=0.2$ jets using the $k_{\rm{T}}$ algorithm~\cite{CATANI1993187,CACCIARI200657} and taking the median jet \pT~density in the event. The two leading $k_{\rm T}$ jets are removed before calculating the median, to suppress the contribution of true hard jets in the background estimation. The correction is applied such that the total momentum removed from the jet is equal to $\rho \times A_{\rm{j}}$, where $A_{\rm{j}}$ is the jet area. This background subtraction is applied both to the measured data and to the embedded PYTHIA reference. 

\section{Observables}

 Jet constituents are reclustered using the physical Cambridge/Aachen (CA) metric \cite{FastJetCA}, leading to an angle-ordered shower.  The declustering process consists of unwinding the clustering history step by step, always following the hardest branch. The first declustering step identifies the final subjet pair or branch that was merged. The second declustering step identifies the subjet pair that was merged into the leading subjet of the final step, etc. 
 The coordinates of the subleading prong in the Lund Plane  (log($z \Delta R$), log(1/$\Delta R$)) are registered at each declustering step, where $z$ is the fraction of momentum carried by the subleading prong $z = \frac{{\rm{min}}(p_{\rm{T,1}},p_{\rm{T,2}})}{p_{\rm{T,1}}+p_{\rm{T,2}}}$,  with $p_{\rm{T,1}}$ and $p_{\rm{T,2}}$  being the momenta of the leading and subleading prongs, respectively, and $\Delta R$ the opening angle of the splitting.  

The observable $n_{\rm{SD}}$ is obtained by counting the number of splittings in the declustering process that satisfy the Soft Drop selection $z > z_{\rm{cut}}$, $z_{\rm{cut}}=0.1$. The observable $z_{\rm g}$ corresponds to the subjet momentum balance, $z$, of the first splitting satisfying the SD selection.
Jets with $n_{\rm{SD}}=0$ are labelled ``untagged jets''. The $z_{\rm g}$ distribution is absolutely normalised, including the untagged jets in the normalisation.  This choice of normalisation, used here for the first time, provides crucial information for quantitative comparison of jet substructure measurements in Pb--Pb and pp collisions since it allows the results to be interpreted in terms of not only a change of shape in the distribution but also in terms of net enhancement/suppression of the yield of splittings satisfying the SD condition in a given jet transverse momentum range. 

The tracking system enables the measurement of subjets with angular separation smaller than 0.1 radians and a scan of the $\zg$ distribution in ranges of $\Delta R$: $\DeltaR<0.1$, $\DeltaR>0.1$ and $\DeltaR>0.2$.

For data from pp collisions, the correction of the detector effects was performed via unfolding. The results are presented in the jet momentum interval of $40<p_{\rm{T,jet}}^{\rm{ch}}<60$ GeV/$\it{c}$,  chosen to balance statistical precision and detector effects.
In Pb--Pb collisions, the results are presented at detector-level, with the uncorrelated background subtracted on average from the jet $p_{\rm{T}}$ and from the substructure observable. The vacuum reference is thus smeared by background fluctuations and instrumental effects. The Pb--Pb results are presented in the jet momentum range of $80<p_{\rm{T,jet}}^{\rm{ch}}<120$ GeV/$\it{c}$, where uncorrelated background is negligible.

\section{Corrections and systematic uncertainties}

For data from pp collisions, the unfolding of instrumental effects is carried out using a four-dimensional response matrix that encodes the smearing of both jet $p_{\rm{T}}^{\rm{ch}}$ and the substructure observable 
(shape$^{\mathrm{part,ch}}$, $p_{\mathrm{T,jet}}^{\mathrm{part,ch}}$, shape$^{\mathrm{det,ch}}$, $p_{\mathrm{T,jet}}^{\mathrm{det,ch}}$), where ``shape'' denotes either $z_{\rm{g}}$ or $n_{\rm{SD}}$.
The upper index ``part'' refers to particle-level  and ``det'' refers to detector-level quantities, obtained from simulations in which pp collisions are generated by PYTHIA (particle-level) and then passed through a GEANT3-based model \cite{GEANT3} of the ALICE detector. We note that the particle-level jet finding is performed using the true particle masses so the unfolding corrects for the pion mass assumption at detector level. 

To generate vacuum reference distributions for comparison to Pb--Pb results, which are not fully corrected, we superimpose detector-level PYTHIA events onto real Pb--Pb events. Consequently, no two-track effects are present, however their impact in data is negligible due to the large required number of clusters per track. The matching of particle-level and embedded jets is performed as described in \cite{Acharya:2018uvf}. The matching efficiency is consistent with unity for jets with $p_{\rm{T}}$ above 30\,GeV/$c$.

For pp collisions, Bayesian unfolding in two dimensions as implemented in the  RooUnfold package~\cite{RooUnfold} is used. 
The prior is the two-dimensional distribution ($p_{\rm{T,jet}}^{\rm{part,ch}}$, shape$^{\mathrm{part,ch}}$) generated with PYTHIA. The default number of iterations chosen for $z_{\rm g}$ and $n_{\rm{SD}}$ is 4, which corresponds to the first iteration for which the refolded distributions agree with the corresponding raw distributions within 5$\%$.  A closure test was also carried out, in which two statistically independent Monte Carlo (MC) samples are used to fill the response and
the pseudo-data. For this test, the unfolded solution agrees with the MC truth distribution within statistical uncertainties.

Unfolding of the distributions was attempted for the Pb--Pb case, but no convergence on a mathematically consistent solution was obtained, due to the limited statistics of the data sample and due to the fact that the response is strongly non-diagonal due to the presence of sub-leading prongs at large angles that are not correlated to particle-level prongs and that arise due to fluctuations of the uncorrelated background. Strategies to suppress such secondary prongs are beyond the scope of this analysis.

\begin{table*}[!t]
\centering
\caption{Relative systematic uncertainties on the measured distributions in pp collisions for three selected jet shape intervals in the jet $p_{\mathrm{T,jet}}^{\rm ch}$ interval of $40$--$60$\,GeV/$c$. Due to the shape of the $n_{\rm{SD}}$ distribution, the systematic variations lead to a crossing at central values which artificially reduces the evaluated systematic uncertainty. To improve this we smooth the total systematic uncertainty by interpolating between neighbouring bins. }
\renewcommand{\arraystretch}{1.3}
\begin{tabular}{|l|ccc|ccc|}
\hline 
\multicolumn{1}{|l|}{Observable} & \multicolumn{3}{c|}{$z_{\rm g}$} 
 & \multicolumn{3}{c|}{$n_{\rm{SD}}$} \\
\hline
\hline
\multicolumn{1}{|l|}{Interval} 
& 0.1--0.175 & 0.25--0.325 & 0.4--0.5 & 0 & 3 & 6 \\
\cline{1-7}

Tracking efficiency ($\%$)             & $1.9$ & $0.2$  & $1.0$ 
                    
                      & $16.1$ &$1.1$  & $18.3$  \\

Prior ($\%$)                  & $_{-3.9}^{+0.0}$ & $_{-0.0}^{+7.6}$  & $_{-9.4}^{+0.0}$ 
                     
                      & $_{-1.6}^{+0.0}$ &
                                                               $_{-9.2}^{+0.0}$  & $_{-21.3}^{+0.0}$  \\

Regularisation ($\%$)   & $_{-0.5}^{+0.8}$ & $_{-0.2}^{+0.2}$  & $_{-0.5}^{+0.4}$ 
                    
                      & $_{-1.4}^{+0.4}$ &
                                                               $_{-1.1}^{+1.4}$  & $_{-3.0}^{+1.7}$  \\

 Truncation ($\%$)   & $_{-0.0}^{+2.2}$ & $_{-0.0}^{+1.8}$  & $_{-0.0}^{+2.4}$ 
                   
                      & $_{-0.0}^{+0.0}$ &
                                                               $_{-0.1}^{+0.0}$  & $_{-0.0}^{+4.4}$  \\

 Binning ($\%$)   & $0.5$ & $4.5$  & $1.2$ 
                  
                      & N/A &
                                                               N/A  & N/A  \\
 \cline{1-7}
 Total ($\%$)   & $_{-4.4}^{+3.0}$ & $_{-3.0}^{+8.2}$  & $_{-9.6}^{+2.6}$
                    
                      & $_{-16.2}^{+16.1}$ & $_{-6.2}^{+7.8}$ & $_{-28.2}^{+18.9}$ \\
                     
\hline
\end{tabular}

\label{tab:Syspp}
\end{table*}

\begin{table*}[!t]
\centering
\caption{Relative systematic uncertainties on the measured distributions in Pb--Pb collisions for three selected jet shape intervals and one $\Delta R$ selection in the jet $p_{\mathrm{T,jet}}^{\rm ch}$ interval of $80$--$120$\,GeV/$c$. }
\renewcommand{\arraystretch}{1.3}
\begin{tabular}{|l|ccc|ccc|}
\hline 
\multicolumn{1}{|l|}{Observable} & \multicolumn{3}{c|}{$z_{\rm g}(\Delta R > 0.1)$} & \multicolumn{3}{c|}{$n_{\rm{SD}}$} \\
\hline
\hline
\multicolumn{1}{|l|}{Interval} 
& 0.1--0.175 & 0.25--0.325 & 0.4--0.5 & 0 & 3 & 6 \\
\cline{1-7}

Tracking efficiency($\%$)            & $4.9$ & $2.8$  & $11.4$ 
                            & $11.2$ &$7.9$  & $11.1$  \\

Angular cutoff ($\%$)   & $_{-3.8}^{+2.3}$ & $_{-0.0}^{+2.8}$  & $_{-0.0}^{+10.0}$ 
                         & N/A &
                                                             N/A  & N/A  \\

 Reference ($\%$)   & $_{-0.0}^{+0.0}$ & $_{-0.0}^{+12.4}$  & $_{-0.0}^{+10.1}$ 
                   & $_{-0.0}^{+30.1}$ &
                                                               $_{-5.2}^{+0.0}$  & $_{-0.0}^{+5.3}$  \\

 \cline{1-7}
 Total ($\%$)   & $_{-6.2}^{+5.4}$ & $_{-2.8}^{+13.1}$  & $_{-11.4}^{+18.2}$
                   
                      & $_{-11.2}^{+32.1}$ & $_{-9.5}^{+7.9}$ & $_{-11.1}^{+12.3}$ \\
                     
\hline
\end{tabular}

\label{tab:SysPbPb}
\end{table*}

The systematic uncertainties are determined by varying key aspects of the correction procedures for instrumental response and background fluctuations. The most significant components of the systematic uncertainties for $z_{\rm g}$ and  $n_{\rm{SD}}$ are tabulated in Table~\ref{tab:Syspp} and \ref{tab:SysPbPb}.
For pp collisions, the tracking efficiency uncertainty is $\pm 4\%$ \cite{Acharya:2017goa}. The effect of this uncertainty on the substructure measurement is assessed by applying an additional track rejection of $4\%$ at detector-level prior to jet finding. A new response is built and the unfolding is repeated, with the resulting variation in the unfolded solution symmetrised and taken as the systematic uncertainty. This is the largest contribution to the JES uncertainty. To estimate the regularisation uncertainty, the number of Bayesian iterations is varied by $\pm1$ with respect to the default analysis value. The prior is varied by reweighting the response such that its particle-level projection (PYTHIA) matches HERWIG 7.1.2 \cite{Bellm:2015jjp}. The detector-level intervals in $p_{\rm T}$ and the substructure observables are modified to determine what in the table is referred to as truncation uncertainty. The uncertainty labelled ``Binning'' in the tables corresponds to a variation in binning of both  $p_{\rm T}$  and substructure observables, subject to the constraint of at least 10 counts in the least populous bin to ensure the stability of the unfolding procedure.

In the case of Pb--Pb collisions, the evaluation of the uncertainty due to tracking efficiency is carried out similarly to the pp case. The $z_{\rm g}$ measurement is done differentially in ranges of  $\Delta R$. The limits of the $\Delta R$ ranges were varied by $\pm 10\%$, which corresponds approximately to the width of the distribution of the relative difference of particle-level and embedded-level $\Delta R$ in Pb--Pb collisions. 
The differences between PYTHIA and the unfolded pp distributions are taken into account when using PYTHIA as a reference for Pb--Pb measurements. This is done by reweighting the embedded PYTHIA reference so that its particle-level projection matches the unfolded pp $p_{\rm{T,jet}}$ vs $z_{\rm g}$ (or $p_{\rm{T,jet}}$ vs $n_{\rm{SD}}$) correlation. The difference between the reference smeared with the default and the reweighted response is assigned as the corresponding uncertainty. 
 
In both the pp and Pb--Pb analyses, the uncertainties are added in quadrature. All the contributions to the overall uncertainty produce changes in a given interval of the distribution that are strongly anti-correlated with changes in a different interval, i.e., they induce changes in the shape of the observable.


\section{Results}

\begin{figure}[!h]
\centering
\includegraphics[width=0.65\textwidth]{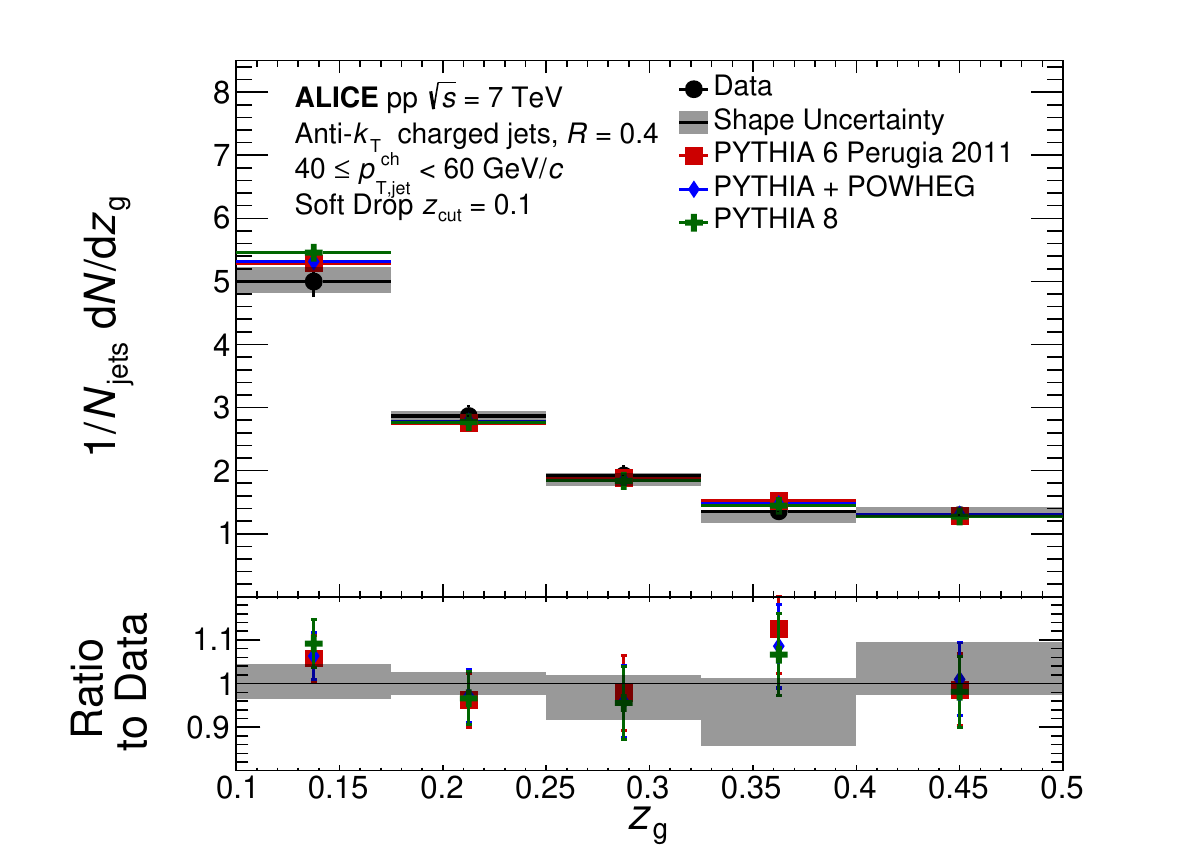}
 \caption{Fully corrected \zg~ distribution in pp collisions for 40 $\leq \pTjetch <$ 60 GeV/{\it c} compared with predictions from PYTHIA simulations. The statistical uncertainties are shown as vertical bars and the systematic uncertainties are represented by a shaded area.}
 \label{fig:CorrectedResultppzg}
 \end{figure}

\begin{figure}[!h]
\centering
\includegraphics[width=0.65\textwidth]{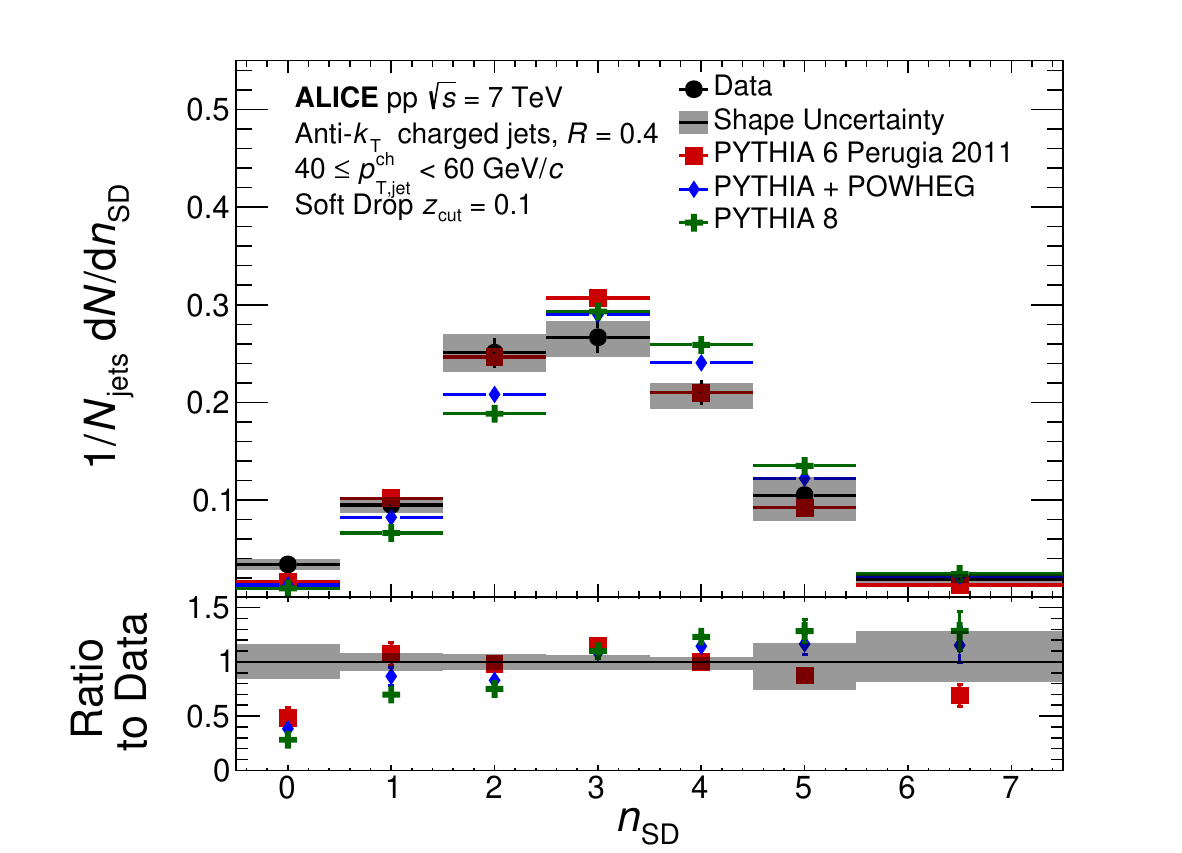}
 \caption{Fully corrected \nsd~distribution in pp collisions for 40 $\leq \pTjetch <$ 60 GeV/{\it c}, compared with predictions from PYTHIA simulations. The statistical uncertainties are shown as vertical bars and the systematic uncertainties are represented by a shaded area.}
 \label{fig:CorrectedResultppnsd}
 \end{figure}

Figures~\ref{fig:CorrectedResultppzg} and \ref{fig:CorrectedResultppnsd} show fully corrected distributions of \zg~and \nsd~measured in pp collisions at $\sqrts = 7$~TeV for charged jets in the interval 40 $< \pTjetch <$ 60 GeV/{\it c}. The results are compared to distributions obtained from PYTHIA 6 (Perugia Tune 2011), from  PYTHIA 6 + POWHEG~\cite{Frixione:2007vw}, to consider the impact of NLO effects, and from the newer  PYTHIA 8 (Tune 4C) \cite{Sjostrand:2014zea}. 

The \zg~distribution is well-described within systematic and statistical uncertainties by all the MC generators considered. As discussed above, untagged jets contribute to the normalisation of the distributions. The untagged contribution is not shown in Fig.~\ref{fig:CorrectedResultppzg}, due to the suppressed zero on the horizontal axis, but is shown in  Fig.~\ref{fig:CorrectedResultppnsd} in the bin representing $n_{\rm{SD}}=0$. Table~\ref{tab:TagRates} shows the tagged fraction for data and simulations. For pp (rightmost column), the untagged fraction is about $2\%$. 
The Monte Carlo distributions in Fig.~\ref{fig:CorrectedResultppnsd} disagree with the data in the tails of the distribution. They have a significantly lower fraction of jets with no splittings (\nsd~$= 0$) than observed in data.
The addition of POWHEG corrections to PYTHIA 6 induces a small shift of the distribution towards a larger number of splittings. 

\begin{figure*}[!t]
\centering
\includegraphics[width=\textwidth]{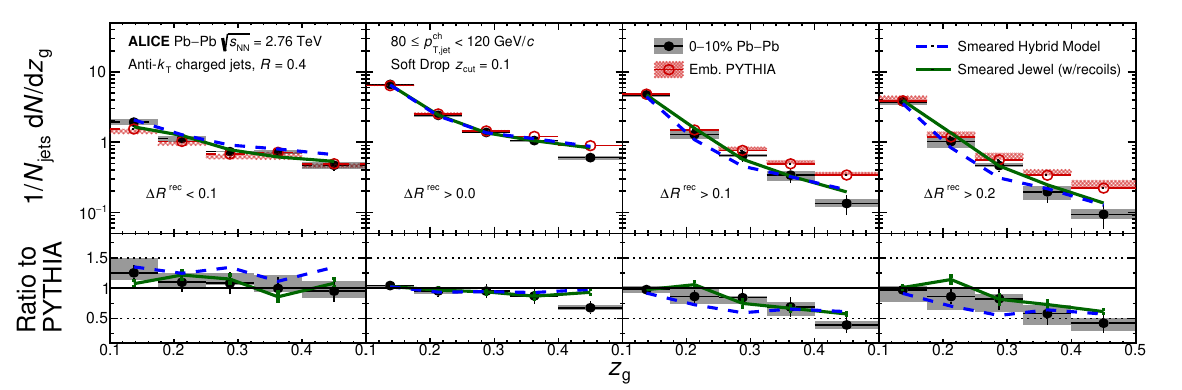}
\caption{Detector-level Pb--Pb distributions of \zg~for $R=0.4$ jets with varying minimum/maximum angular separation of subjets ($\DeltaR$) for jets in the range 80 $\leq \pTjetch <$ 120\, GeV/$c$. The systematic uncertainties are represented by the shaded area. The corresponding values for the embedded PYTHIA reference (open symbols), Hybrid model (dashed line) and JEWEL (solid line) are also shown in the plot. The lower plots show the ratios of data, Hybrid and JEWEL model to the embedded PYTHIA reference.  }
\label{fig:PbPbRawzgDiffDelR}
\end{figure*}

Figure~\ref{fig:PbPbRawzgDiffDelR} shows \zg~distributions measured in central Pb--Pb collisions for various ranges of angular separation \DeltaR. The results are presented in the uncorrected transverse momentum range $80  \leq \pTjetch <  120$ \gev~and compared to the distribution of PYTHIA jets embedded into real 0--10\% central Pb--Pb events.  

Figure~\ref{fig:PbPbRawzgDiffDelR} shows a larger difference between the measured Pb--Pb and embedded reference distributions for larger values of $\Delta{R}$, indicating a relative suppression in the rate of symmetric splittings ($z_{\rm g} \approx 0.5$) in central Pb--Pb collisions. However, due to the steeply falling $z_{\rm{g}}$ distribution, the fraction of all jets that exhibit symmetric splittings is small, and this strong suppression corresponds to a suppression of only a few percent in the total rate of jets passing both the SD and angular cuts (c.f. Tab. \ref{tab:TagRates}). Conversely, in the small $\Delta R$ limit a small excess of splittings is observed in the data.

Figure~\ref{fig:PbPbRawzgDiffDelR} also shows comparisons to predictions from the JEWEL event generator~\cite{Zapp:2008gi} and Hybrid model~\cite{Casalderrey-Solana:2014wca} calculations. The JEWEL simulations include the medium response from jet-medium interactions~\cite{KunnawalkamElayavalli:2017hxo}. The theoretical predictions must be smeared to account for the detector effects as well as fluctuations due to uncorrelated background. This smearing is performed by constructing a 6-dimensional response matrix by superimposing PYTHIA events at detector level to real 0--10$\%$ central Pb--Pb events. The 6-dimensional matrix maps every embedded jet from a given bin of ($\zg^{\rm{part}}$, $\DeltaR^{\rm{part}}$, $p_{\rm{T,jet}}^{\rm{part}}$) to ($\zg^{\rm{det}}$, $\DeltaR^{\rm{det}}$, $p_{\rm{T,jet}}^{\rm{det}}$). The smearing of the distributions significantly modifies the predictions and is essential for quantitative comparison of the measurements and calculations.

The models capture the qualitative trends of the data, namely the enhancement of the number of small-angle splittings and the suppression of the large-angle symmetric splittings. The fraction of jets not passing the SD selection is similar in the models and data. However discrepancies are observed in the angular selection. For instance the number of SD splittings that pass the angular cut of $\Delta R>0.2$ is the lowest in the case of the Hybrid model, pointing to a stronger incoherent quenching of the prongs. 

The suppression of splittings at large opening angles is qualitatively expected from vacuum formation time and colour coherence arguments~\cite{Mehtar-Tani:2016aco}. The wider the opening angle, the shorter the formation time of the splitting. This makes it more likely that the splitting propagates through, and is modified by, the medium. If coherence effects are at play in the medium then it is expected that splittings that are separated by more than the coherence angle will be more suppressed since they radiate energy independently. 

\begin{table*}[!t]
\centering
\caption{Fraction of jets that pass the Soft Drop condition $\zcut = 0.1$ in the specified range of angular separation and in the transverse momentum range $40 \leq \pTjetch < 60\,  \gev$ for pp and $80 \leq \pTjetch < 120$ GeV/$c$ for Pb--Pb collisions. Uncertainties on the data are written as statistical (systematic).}
\setlength\tabcolsep{2pt}
\renewcommand{\arraystretch}{1.3}
\begin{tabular}{|||c|c|c|c|c|c|}
\hline 
\multicolumn{1}{|l|}{} & \multicolumn{5}{c|}{Tagged rate ($\%$)}  \\
\hline
\multicolumn{1}{|c|}{Dataset} & \multicolumn{4}{c|}{Pb--Pb} & \multicolumn{1}{c|}{pp} \\
\hline
\multicolumn{1}{|l|}{Angular Cut} & $\DeltaR < 0.1$ & $\DeltaR > 0.0$  & $\DeltaR > 0.1$ & $\DeltaR > 0.2$ & $\DeltaR > 0.0$  \\
\hline
\multicolumn{1}{|l|}{Data} & $38.4\pm 2.3(2.5)$ & $92.1\pm 3.5(0.9)$  & $53.6\pm 2.7(3.4)$ & $41.8\pm 2.4(3.6)$ & $97.3\pm 3.0(1.7)$  \\
\multicolumn{1}{|l|}{PYTHIA} & $34.6 $ & $95.5 $  & $60.2 $ & $46.9 $ & $98.6 $  \\
\multicolumn{1}{|l|}{Hybrid} & $47.5 $ & $93.4 $  & $45.8 $ & $35.0 $ & N/A \\
\multicolumn{1}{|l|}{JEWEL} & $42.0 $ & $93.0 $ & $51.0 $ & $40.0 $ & N/A  \\

\hline

\end{tabular}

\label{tab:TagRates}
\end{table*}

\begin{figure}[!h]
\centering
\includegraphics[width=0.65\textwidth]{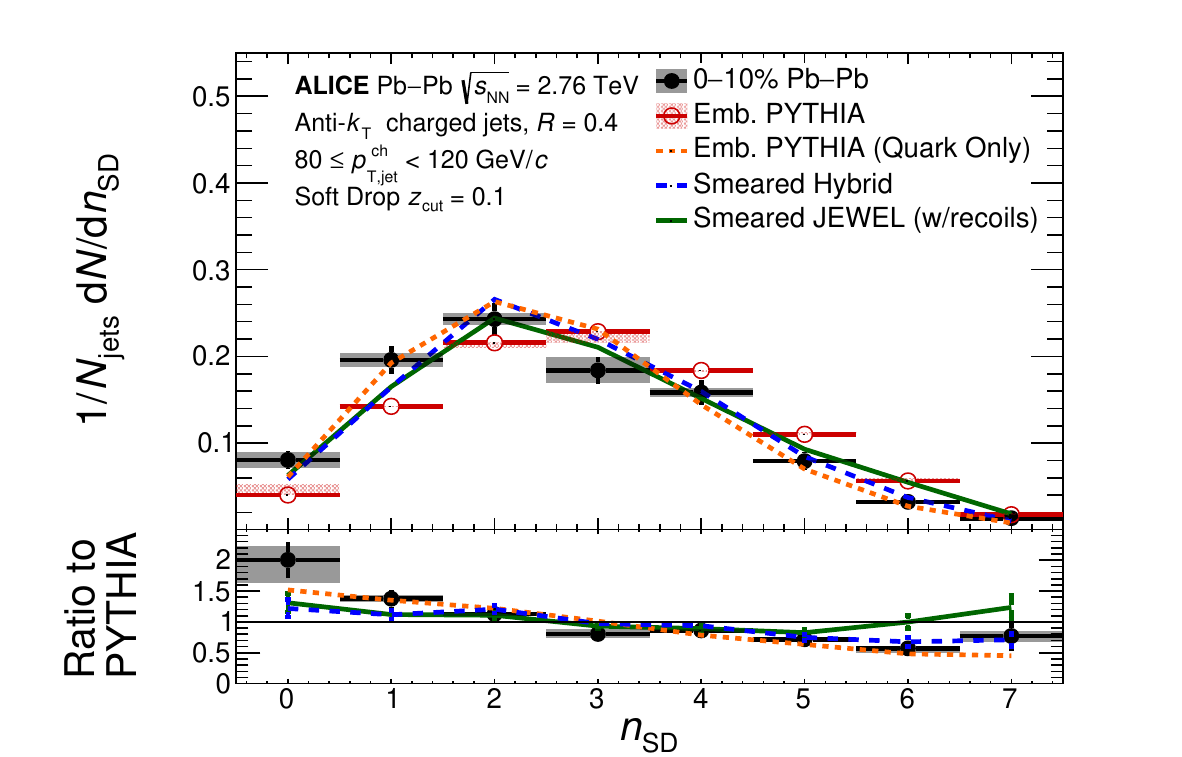}
\caption{The number of SD branches for jets reconstructed in Pb--Pb data are shown. The systematic uncertainties are represented by the shaded area.  The datapoints are compared to jets found in PYTHIA events embedded into Pb--Pb events (open markers). The Hybrid model and JEWEL predictions correspond to the red (dashed) and blue (solid) lines. The lower panel shows the ratio of the $n_{\rm{SD}}$ distribution in data and the embedded PYTHIA reference (grey). The ratios of the Hybrid and JEWEL models to the embedded PYTHIA reference are also shown and their uncertainties are purely statistical}
\label{fig:RecursiveDataRatioSD}
\end{figure}

Figure~\ref{fig:RecursiveDataRatioSD} shows the comparison of $n_{\rm{SD}}$ distributions from Pb--Pb measurements and the embedded PYTHIA reference. The data exhibit a shift towards lower number of splittings. The discrepancies between the distributions from PYTHIA and from pp collisions are incorporated as a part of the reference uncertainty  via the reweighting procedure described above. The corresponding curves for the Hybrid model and JEWEL are also shown in the plot. 

To explore the dependence of the $n_{\rm{SD}}$ distribution on the fragmentation pattern, we also show a calculation in which the pp reference distribution is based solely on light-quark fragmentation. Since the quark fragmentation is harder, we see that the number of splittings peaks at lower values, in line with what we observe in the data. The smeared JEWEL and Hybrid model calculation agree with the qualitative trend of the data. 

 The trends indicate that the larger the opening angle, the more suppressed the splittings are, and this is qualitatively consistent with large-angle prongs being more resolved by the medium and thus more suppressed. The same process could lead to a reduction in the number of hard splittings as observed in Figure ~\ref{fig:RecursiveDataRatioSD}.
 However, it is worth noting that both the Hybrid and JEWEL models, in spite of their capturing of the general trends of the data, they do not incorporate the physics of color coherence and all the prongs in the jet lose energy incoherently. This points to a simpler interpretation of the results for instance in terms of formation times of the splittings and their interplay with the medium length. The vacuum formation time $t_{\rm f }\approx \omega/k_{\rm T}^{2} \approx 1/(\omega \Delta R^{2})$, with $\omega$ and $k_{\rm T}$ being the energy and relative transverse momentum of the radiated prong, is shorter for large-angle splittings, meaning that vacuum, large-angle splittings, will be produced mostly in the medium and their resulting prongs will be further modified by the medium. At large angles, the component of vacuum splittings that propagate in vacuum is less than at small angles, resulting in an enhanced contribution of medium-modifications compared to small-angle splittings.

\section{Summary}

This Letter presents the measurement of jet substructure using iterative declustering techniques in pp and Pb--Pb collisions at the LHC.
We report distributions of $n_{\rm{SD}}$, the number of branches passing the soft drop selection, and $z_{\rm g}$, the shared momentum fraction of the two-prong substructure selected by the mass drop condition, differentially in ranges of splitting opening angle.  

Generally, good agreement between distributions for pp collisions and vacuum calculations is found except for the fraction of untagged jets, which is underestimated by the models. In Pb--Pb collisions, a suppression of the \zg~distribution is observed at large angles relative to the vacuum reference whilst at low opening angles there is a hint of an enhancement. These observations are in qualitative agreement with the expected behaviour of two-prong objects in the case of coherent or decoherent energy loss \cite{Mehtar-Tani:2016aco} in the BMDPS-Z~\cite{Baier:1994tc,Baier:2000mf} framework. 
However, the models that are compared to the data do not implement color coherence and yet they capture the qualitative trends of the data. This suggests that other effects might drive the observed behaviour, for instance the interplay between formation time of the splittings and medium length.

The number of splittings obtained by iteratively declustering the hardest branch in the jet, \nsd, is shifted towards lower values in Pb--Pb relative to the vacuum reference.  This suggests that  medium-induced radiation does not create new splittings that pass the SD cut. On the contrary, there is a hint of fewer splittings passing the SD cut, pointing to a harder, more quark-like fragmentation in Pb--Pb compared to pp collisions, in qualitative agreement with the trends observed for other jet shapes \cite{Acharya:2018uvf}.    

With these measurements, we have explored a region of the Lund plane delimited by the Soft Drop cut $z>0.1$. Other regions of the phase space of splittings will be scanned systematically in the future with larger data samples.


\newenvironment{acknowledgement}{\relax}{\relax}
\begin{acknowledgement}
\section*{Acknowledgements}

The ALICE Collaboration would like to thank all its engineers and technicians for their invaluable contributions to the construction of the experiment and the CERN accelerator teams for the outstanding performance of the LHC complex.
The ALICE Collaboration gratefully acknowledges the resources and support provided by all Grid centres and the Worldwide LHC Computing Grid (WLCG) collaboration.
The ALICE Collaboration acknowledges the following funding agencies for their support in building and running the ALICE detector:
A. I. Alikhanyan National Science Laboratory (Yerevan Physics Institute) Foundation (ANSL), State Committee of Science and World Federation of Scientists (WFS), Armenia;
Austrian Academy of Sciences, Austrian Science Fund (FWF): [M 2467-N36] and Nationalstiftung f\"{u}r Forschung, Technologie und Entwicklung, Austria;
Ministry of Communications and High Technologies, National Nuclear Research Center, Azerbaijan;
Conselho Nacional de Desenvolvimento Cient\'{\i}fico e Tecnol\'{o}gico (CNPq), Universidade Federal do Rio Grande do Sul (UFRGS), Financiadora de Estudos e Projetos (Finep) and Funda\c{c}\~{a}o de Amparo \`{a} Pesquisa do Estado de S\~{a}o Paulo (FAPESP), Brazil;
Ministry of Science \& Technology of China (MSTC), National Natural Science Foundation of China (NSFC) and Ministry of Education of China (MOEC) , China;
Croatian Science Foundation and Ministry of Science and Education, Croatia;
Centro de Aplicaciones Tecnol\'{o}gicas y Desarrollo Nuclear (CEADEN), Cubaenerg\'{\i}a, Cuba;
Ministry of Education, Youth and Sports of the Czech Republic, Czech Republic;
The Danish Council for Independent Research | Natural Sciences, the Carlsberg Foundation and Danish National Research Foundation (DNRF), Denmark;
Helsinki Institute of Physics (HIP), Finland;
Commissariat \`{a} l'Energie Atomique (CEA), Institut National de Physique Nucl\'{e}aire et de Physique des Particules (IN2P3) and Centre National de la Recherche Scientifique (CNRS) and R\'{e}gion des  Pays de la Loire, France;
Bundesministerium f\"{u}r Bildung und Forschung (BMBF) and GSI Helmholtzzentrum f\"{u}r Schwerionenforschung GmbH, Germany;
General Secretariat for Research and Technology, Ministry of Education, Research and Religions, Greece;
National Research, Development and Innovation Office, Hungary;
Department of Atomic Energy Government of India (DAE), Department of Science and Technology, Government of India (DST), University Grants Commission, Government of India (UGC) and Council of Scientific and Industrial Research (CSIR), India;
Indonesian Institute of Science, Indonesia;
Centro Fermi - Museo Storico della Fisica e Centro Studi e Ricerche Enrico Fermi and Istituto Nazionale di Fisica Nucleare (INFN), Italy;
Institute for Innovative Science and Technology , Nagasaki Institute of Applied Science (IIST), Japan Society for the Promotion of Science (JSPS) KAKENHI and Japanese Ministry of Education, Culture, Sports, Science and Technology (MEXT), Japan;
Consejo Nacional de Ciencia (CONACYT) y Tecnolog\'{i}a, through Fondo de Cooperaci\'{o}n Internacional en Ciencia y Tecnolog\'{i}a (FONCICYT) and Direcci\'{o}n General de Asuntos del Personal Academico (DGAPA), Mexico;
Nederlandse Organisatie voor Wetenschappelijk Onderzoek (NWO), Netherlands;
The Research Council of Norway, Norway;
Commission on Science and Technology for Sustainable Development in the South (COMSATS), Pakistan;
Pontificia Universidad Cat\'{o}lica del Per\'{u}, Peru;
Ministry of Science and Higher Education and National Science Centre, Poland;
Korea Institute of Science and Technology Information and National Research Foundation of Korea (NRF), Republic of Korea;
Ministry of Education and Scientific Research, Institute of Atomic Physics and Ministry of Research and Innovation and Institute of Atomic Physics, Romania;
Joint Institute for Nuclear Research (JINR), Ministry of Education and Science of the Russian Federation, National Research Centre Kurchatov Institute, Russian Science Foundation and Russian Foundation for Basic Research, Russia;
Ministry of Education, Science, Research and Sport of the Slovak Republic, Slovakia;
National Research Foundation of South Africa, South Africa;
Swedish Research Council (VR) and Knut \& Alice Wallenberg Foundation (KAW), Sweden;
European Organization for Nuclear Research, Switzerland;
National Science and Technology Development Agency (NSDTA), Suranaree University of Technology (SUT) and Office of the Higher Education Commission under NRU project of Thailand, Thailand;
Turkish Atomic Energy Agency (TAEK), Turkey;
National Academy of  Sciences of Ukraine, Ukraine;
Science and Technology Facilities Council (STFC), United Kingdom;
National Science Foundation of the United States of America (NSF) and United States Department of Energy, Office of Nuclear Physics (DOE NP), United States of America.
\end{acknowledgement}

\bibliographystyle{utphys}   
\bibliography{references}

\newpage
\appendix

%
%

\section{The ALICE Collaboration}
\label{app:collab}

\begingroup
\small
\begin{flushleft}
S.~Acharya\Irefn{org141}\And 
D.~Adamov\'{a}\Irefn{org93}\And 
S.P.~Adhya\Irefn{org141}\And 
A.~Adler\Irefn{org74}\And 
J.~Adolfsson\Irefn{org80}\And 
M.M.~Aggarwal\Irefn{org98}\And 
G.~Aglieri Rinella\Irefn{org34}\And 
M.~Agnello\Irefn{org31}\And 
N.~Agrawal\Irefn{org48}\textsuperscript{,}\Irefn{org10}\And 
Z.~Ahammed\Irefn{org141}
% ...
\And
\renewcommand\labelenumi{\textsuperscript{\theenumi}~}

\section*{Affiliation notes}
\renewcommand\theenumi{\roman{enumi}}
\begin{Authlist}
\item \Adef{org*}Deceased
\item \Adef{orgI}Dipartimento DET del Politecnico di Torino, Turin, Italy
\item \Adef{orgII}M.V. Lomonosov Moscow State University, D.V. Skobeltsyn Institute of Nuclear, Physics, Moscow, Russia
\item \Adef{orgIII}Department of Applied Physics, Aligarh Muslim University, Aligarh, India
\item \Adef{orgIV}Institute of Theoretical Physics, University of Wroclaw, Poland
\end{Authlist}

\section*{Collaboration Institutes}
\renewcommand\theenumi{\arabic{enumi}~}
\begin{Authlist}
\item \Idef{org1}A.I. Alikhanyan National Science Laboratory (Yerevan Physics Institute) Foundation, Yerevan, Armenia
\item \Idef{org2}Bogolyubov Institute for Theoretical Physics, National Academy of Sciences of Ukraine, Kiev, Ukraine
\item \Idef{org3}Bose Institute, Department of Physics  and Centre for Astroparticle Physics and Space Science (CAPSS), Kolkata, India
\item \Idef{org4}Budker Institute for Nuclear Physics, Novosibirsk, Russia
\item \Idef{org5}California Polytechnic State University, San Luis Obispo, California, United States
\item \Idef{org6}Central China Normal University, Wuhan, China
\item \Idef{org7}Centre de Calcul de l'IN2P3, Villeurbanne, Lyon, France
\item \Idef{org8}Centro de Aplicaciones Tecnol\'{o}gicas y Desarrollo Nuclear (CEADEN), Havana, Cuba
\item \Idef{org9}Centro de Investigaci\'{o}n y de Estudios Avanzados (CINVESTAV), Mexico City and M\'{e}rida, Mexico
\item \Idef{org10}Centro Fermi - Museo Storico della Fisica e Centro Studi e Ricerche ``Enrico Fermi', Rome, Italy
\item \Idef{org11}Chicago State University, Chicago, Illinois, United States
\item \Idef{org12}China Institute of Atomic Energy, Beijing, China
\item \Idef{org13}Chonbuk National University, Jeonju, Republic of Korea
\item \Idef{org14}Comenius University Bratislava, Faculty of Mathematics, Physics and Informatics, Bratislava, Slovakia
\item \Idef{org15}COMSATS University Islamabad, Islamabad, Pakistan
\item \Idef{org16}Creighton University, Omaha, Nebraska, United States
\item \Idef{org17}Department of Physics, Aligarh Muslim University, Aligarh, India
\item \Idef{org18}Department of Physics, Pusan National University, Pusan, Republic of Korea
\item \Idef{org19}Department of Physics, Sejong University, Seoul, Republic of Korea
\item \Idef{org20}Department of Physics, University of California, Berkeley, California, United States
\item \Idef{org21}Department of Physics, University of Oslo, Oslo, Norway
\item \Idef{org22}Department of Physics and Technology, University of Bergen, Bergen, Norway
\item \Idef{org23}Dipartimento di Fisica dell'Universit\`{a} 'La Sapienza' and Sezione INFN, Rome, Italy
\item \Idef{org24}Dipartimento di Fisica dell'Universit\`{a} and Sezione INFN, Cagliari, Italy
\item \Idef{org25}Dipartimento di Fisica dell'Universit\`{a} and Sezione INFN, Trieste, Italy
\item \Idef{org26}Dipartimento di Fisica dell'Universit\`{a} and Sezione INFN, Turin, Italy
\item \Idef{org27}Dipartimento di Fisica e Astronomia dell'Universit\`{a} and Sezione INFN, Bologna, Italy
\item \Idef{org28}Dipartimento di Fisica e Astronomia dell'Universit\`{a} and Sezione INFN, Catania, Italy
\item \Idef{org29}Dipartimento di Fisica e Astronomia dell'Universit\`{a} and Sezione INFN, Padova, Italy
\item \Idef{org30}Dipartimento di Fisica `E.R.~Caianiello' dell'Universit\`{a} and Gruppo Collegato INFN, Salerno, Italy
\item \Idef{org31}Dipartimento DISAT del Politecnico and Sezione INFN, Turin, Italy
\item \Idef{org32}Dipartimento di Scienze e Innovazione Tecnologica dell'Universit\`{a} del Piemonte Orientale and INFN Sezione di Torino, Alessandria, Italy
\item \Idef{org33}Dipartimento Interateneo di Fisica `M.~Merlin' and Sezione INFN, Bari, Italy
\item \Idef{org34}European Organization for Nuclear Research (CERN), Geneva, Switzerland
\item \Idef{org35}Faculty of Electrical Engineering, Mechanical Engineering and Naval Architecture, University of Split, Split, Croatia
\item \Idef{org36}Faculty of Engineering and Science, Western Norway University of Applied Sciences, Bergen, Norway
\item \Idef{org37}Faculty of Nuclear Sciences and Physical Engineering, Czech Technical University in Prague, Prague, Czech Republic
\item \Idef{org38}Faculty of Science, P.J.~\v{S}af\'{a}rik University, Ko\v{s}ice, Slovakia
\item \Idef{org39}Frankfurt Institute for Advanced Studies, Johann Wolfgang Goethe-Universit\"{a}t Frankfurt, Frankfurt, Germany
\item \Idef{org40}Gangneung-Wonju National University, Gangneung, Republic of Korea
\item \Idef{org41}Gauhati University, Department of Physics, Guwahati, India
\item \Idef{org42}Helmholtz-Institut f\"{u}r Strahlen- und Kernphysik, Rheinische Friedrich-Wilhelms-Universit\"{a}t Bonn, Bonn, Germany
\item \Idef{org43}Helsinki Institute of Physics (HIP), Helsinki, Finland
\item \Idef{org44}High Energy Physics Group,  Universidad Aut\'{o}noma de Puebla, Puebla, Mexico
\item \Idef{org45}Hiroshima University, Hiroshima, Japan
\item \Idef{org46}Hochschule Worms, Zentrum  f\"{u}r Technologietransfer und Telekommunikation (ZTT), Worms, Germany
\item \Idef{org47}Horia Hulubei National Institute of Physics and Nuclear Engineering, Bucharest, Romania
\item \Idef{org48}Indian Institute of Technology Bombay (IIT), Mumbai, India
\item \Idef{org49}Indian Institute of Technology Indore, Indore, India
\item \Idef{org50}Indonesian Institute of Sciences, Jakarta, Indonesia
\item \Idef{org51}INFN, Laboratori Nazionali di Frascati, Frascati, Italy
\item \Idef{org52}INFN, Sezione di Bari, Bari, Italy
\item \Idef{org53}INFN, Sezione di Bologna, Bologna, Italy
\item \Idef{org54}INFN, Sezione di Cagliari, Cagliari, Italy
\item \Idef{org55}INFN, Sezione di Catania, Catania, Italy
\item \Idef{org56}INFN, Sezione di Padova, Padova, Italy
\item \Idef{org57}INFN, Sezione di Roma, Rome, Italy
\item \Idef{org58}INFN, Sezione di Torino, Turin, Italy
\item \Idef{org59}INFN, Sezione di Trieste, Trieste, Italy
\item \Idef{org60}Inha University, Incheon, Republic of Korea
\item \Idef{org61}Institut de Physique Nucl\'{e}aire d'Orsay (IPNO), Institut National de Physique Nucl\'{e}aire et de Physique des Particules (IN2P3/CNRS), Universit\'{e} de Paris-Sud, Universit\'{e} Paris-Saclay, Orsay, France
\item \Idef{org62}Institute for Nuclear Research, Academy of Sciences, Moscow, Russia
\item \Idef{org63}Institute for Subatomic Physics, Utrecht University/Nikhef, Utrecht, Netherlands
\item \Idef{org64}Institute for Theoretical and Experimental Physics, Moscow, Russia
\item \Idef{org65}Institute of Experimental Physics, Slovak Academy of Sciences, Ko\v{s}ice, Slovakia
\item \Idef{org66}Institute of Physics, Homi Bhabha National Institute, Bhubaneswar, India
\item \Idef{org67}Institute of Physics of the Czech Academy of Sciences, Prague, Czech Republic
\item \Idef{org68}Institute of Space Science (ISS), Bucharest, Romania
\item \Idef{org69}Institut f\"{u}r Kernphysik, Johann Wolfgang Goethe-Universit\"{a}t Frankfurt, Frankfurt, Germany
\item \Idef{org70}Instituto de Ciencias Nucleares, Universidad Nacional Aut\'{o}noma de M\'{e}xico, Mexico City, Mexico
\item \Idef{org71}Instituto de F\'{i}sica, Universidade Federal do Rio Grande do Sul (UFRGS), Porto Alegre, Brazil
\item \Idef{org72}Instituto de F\'{\i}sica, Universidad Nacional Aut\'{o}noma de M\'{e}xico, Mexico City, Mexico
\item \Idef{org73}iThemba LABS, National Research Foundation, Somerset West, South Africa
\item \Idef{org74}Johann-Wolfgang-Goethe Universit\"{a}t Frankfurt Institut f\"{u}r Informatik, Fachbereich Informatik und Mathematik, Frankfurt, Germany
\item \Idef{org75}Joint Institute for Nuclear Research (JINR), Dubna, Russia
\item \Idef{org76}Korea Institute of Science and Technology Information, Daejeon, Republic of Korea
\item \Idef{org77}KTO Karatay University, Konya, Turkey
\item \Idef{org78}Laboratoire de Physique Subatomique et de Cosmologie, Universit\'{e} Grenoble-Alpes, CNRS-IN2P3, Grenoble, France
\item \Idef{org79}Lawrence Berkeley National Laboratory, Berkeley, California, United States
\item \Idef{org80}Lund University Department of Physics, Division of Particle Physics, Lund, Sweden
\item \Idef{org81}Nagasaki Institute of Applied Science, Nagasaki, Japan
\item \Idef{org82}Nara Women{'}s University (NWU), Nara, Japan
\item \Idef{org83}National and Kapodistrian University of Athens, School of Science, Department of Physics , Athens, Greece
\item \Idef{org84}National Centre for Nuclear Research, Warsaw, Poland
\item \Idef{org85}National Institute of Science Education and Research, Homi Bhabha National Institute, Jatni, India
\item \Idef{org86}National Nuclear Research Center, Baku, Azerbaijan
\item \Idef{org87}National Research Centre Kurchatov Institute, Moscow, Russia
\item \Idef{org88}Niels Bohr Institute, University of Copenhagen, Copenhagen, Denmark
\item \Idef{org89}Nikhef, National institute for subatomic physics, Amsterdam, Netherlands
\item \Idef{org90}NRC Kurchatov Institute IHEP, Protvino, Russia
\item \Idef{org91}NRNU Moscow Engineering Physics Institute, Moscow, Russia
\item \Idef{org92}Nuclear Physics Group, STFC Daresbury Laboratory, Daresbury, United Kingdom
\item \Idef{org93}Nuclear Physics Institute of the Czech Academy of Sciences, \v{R}e\v{z} u Prahy, Czech Republic
\item \Idef{org94}Oak Ridge National Laboratory, Oak Ridge, Tennessee, United States
\item \Idef{org95}Ohio State University, Columbus, Ohio, United States
\item \Idef{org96}Petersburg Nuclear Physics Institute, Gatchina, Russia
\item \Idef{org97}Physics department, Faculty of science, University of Zagreb, Zagreb, Croatia
\item \Idef{org98}Physics Department, Panjab University, Chandigarh, India
\item \Idef{org99}Physics Department, University of Jammu, Jammu, India
\item \Idef{org100}Physics Department, University of Rajasthan, Jaipur, India
\item \Idef{org101}Physikalisches Institut, Eberhard-Karls-Universit\"{a}t T\"{u}bingen, T\"{u}bingen, Germany
\item \Idef{org102}Physikalisches Institut, Ruprecht-Karls-Universit\"{a}t Heidelberg, Heidelberg, Germany
\item \Idef{org103}Physik Department, Technische Universit\"{a}t M\"{u}nchen, Munich, Germany
\item \Idef{org104}Politecnico di Bari, Bari, Italy
\item \Idef{org105}Research Division and ExtreMe Matter Institute EMMI, GSI Helmholtzzentrum f\"ur Schwerionenforschung GmbH, Darmstadt, Germany
\item \Idef{org106}Rudjer Bo\v{s}kovi\'{c} Institute, Zagreb, Croatia
\item \Idef{org107}Russian Federal Nuclear Center (VNIIEF), Sarov, Russia
\item \Idef{org108}Saha Institute of Nuclear Physics, Homi Bhabha National Institute, Kolkata, India
\item \Idef{org109}School of Physics and Astronomy, University of Birmingham, Birmingham, United Kingdom
\item \Idef{org110}Secci\'{o}n F\'{\i}sica, Departamento de Ciencias, Pontificia Universidad Cat\'{o}lica del Per\'{u}, Lima, Peru
\item \Idef{org111}Shanghai Institute of Applied Physics, Shanghai, China
\item \Idef{org112}St. Petersburg State University, St. Petersburg, Russia
\item \Idef{org113}Stefan Meyer Institut f\"{u}r Subatomare Physik (SMI), Vienna, Austria
\item \Idef{org114}SUBATECH, IMT Atlantique, Universit\'{e} de Nantes, CNRS-IN2P3, Nantes, France
\item \Idef{org115}Suranaree University of Technology, Nakhon Ratchasima, Thailand
\item \Idef{org116}Technical University of Ko\v{s}ice, Ko\v{s}ice, Slovakia
\item \Idef{org117}Technische Universit\"{a}t M\"{u}nchen, Excellence Cluster 'Universe', Munich, Germany
\item \Idef{org118}The Henryk Niewodniczanski Institute of Nuclear Physics, Polish Academy of Sciences, Cracow, Poland
\item \Idef{org119}The University of Texas at Austin, Austin, Texas, United States
\item \Idef{org120}Universidad Aut\'{o}noma de Sinaloa, Culiac\'{a}n, Mexico
\item \Idef{org121}Universidade de S\~{a}o Paulo (USP), S\~{a}o Paulo, Brazil
\item \Idef{org122}Universidade Estadual de Campinas (UNICAMP), Campinas, Brazil
\item \Idef{org123}Universidade Federal do ABC, Santo Andre, Brazil
\item \Idef{org124}University College of Southeast Norway, Tonsberg, Norway
\item \Idef{org125}University of Cape Town, Cape Town, South Africa
\item \Idef{org126}University of Houston, Houston, Texas, United States
\item \Idef{org127}University of Jyv\"{a}skyl\"{a}, Jyv\"{a}skyl\"{a}, Finland
\item \Idef{org128}University of Liverpool, Liverpool, United Kingdom
\item \Idef{org129}University of Science and Techonology of China, Hefei, China
\item \Idef{org130}University of Tennessee, Knoxville, Tennessee, United States
\item \Idef{org131}University of the Witwatersrand, Johannesburg, South Africa
\item \Idef{org132}University of Tokyo, Tokyo, Japan
\item \Idef{org133}University of Tsukuba, Tsukuba, Japan
\item \Idef{org134}Universit\'{e} Clermont Auvergne, CNRS/IN2P3, LPC, Clermont-Ferrand, France
\item \Idef{org135}Universit\'{e} de Lyon, Universit\'{e} Lyon 1, CNRS/IN2P3, IPN-Lyon, Villeurbanne, Lyon, France
\item \Idef{org136}Universit\'{e} de Strasbourg, CNRS, IPHC UMR 7178, F-67000 Strasbourg, France, Strasbourg, France
\item \Idef{org137}Universit\'{e} Paris-Saclay Centre d'Etudes de Saclay (CEA), IRFU, D\'{e}partment de Physique Nucl\'{e}aire (DPhN), Saclay, France
\item \Idef{org138}Universit\`{a} degli Studi di Foggia, Foggia, Italy
\item \Idef{org139}Universit\`{a} degli Studi di Pavia, Pavia, Italy
\item \Idef{org140}Universit\`{a} di Brescia, Brescia, Italy
\item \Idef{org141}Variable Energy Cyclotron Centre, Homi Bhabha National Institute, Kolkata, India
\item \Idef{org142}Warsaw University of Technology, Warsaw, Poland
\item \Idef{org143}Wayne State University, Detroit, Michigan, United States
\item \Idef{org144}Westf\"{a}lische Wilhelms-Universit\"{a}t M\"{u}nster, Institut f\"{u}r Kernphysik, M\"{u}nster, Germany
\item \Idef{org145}Wigner Research Centre for Physics, Hungarian Academy of Sciences, Budapest, Hungary
\item \Idef{org146}Yale University, New Haven, Connecticut, United States
\item \Idef{org147}Yonsei University, Seoul, Republic of Korea
\end{Authlist}
\endgroup
  
\end{document}